%% file: main.tex
\PassOptionsToPackage{prologue,table}{xcolor}
\documentclass[sigconf]{acmart}

\AtBeginDocument{%
  \providecommand\BibTeX{{%
    \normalfont B\kern-0.5em{\scshape i\kern-0.25em b}\kern-0.8em\TeX}}}

\setcopyright{none}
\renewcommand\footnotetextcopyrightpermission[1]{} 

\input{macro}

\begin{document}


\title{Chatting with GPT-3 for Zero-Shot Human-Like Mobile Automated GUI Testing}

\author{Zhe Liu$^{1,3}$,Chunyang Chen$^4$, Junjie Wang$^{1,2,3,*}$, Mengzhuo Chen$^{1,3}$, Boyu Wu$^{3}$, Xing Che$^{1,3}$, Dandan Wang$^{1,3}$, Qing Wang$^{1,2,3,*}$}
\affiliation{
  \position{$^1$Laboratory for Internet Software Technologies, $^2$State Key Laboratory of Computer Sciences, Science \& Technology on Integrated Information System Laboratory}
  \department{Institute of Software Chinese Academy of Sciences, Beijing, China; \\
  $^3$University of Chinese Academy of Sciences, Beijing, China; $^*$Corresponding author\\
  $^4$Monash University, Melbourne, Australia;
  }
}
\email{liuzhe181@mails.ucas.ac.cn, Chunyang.chen@monash.edu, junjie@iscas.ac.cn, wq@iscas.ac.cn}

\begin{abstract}

Mobile apps are indispensable for people's daily life, and automated GUI (Graphical User
Interface) testing is widely used for app quality assurance.
There is a growing interest in using learning-based techniques for automated GUI testing which aims at generating human-like actions and interactions. 
However, the limitations such as low testing coverage, weak generalization, and heavy reliance on training data, make an urgent need for a more effective approach to generate human-like actions to thoroughly test mobile apps. 
Inspired by the success of the Large Language Model (LLM), e.g., GPT-3 and ChatGPT, in natural language understanding and question answering, we formulate the mobile GUI testing problem as a Q\&A task.
We propose {\tool}, asking LLM to chat with the mobile apps by passing the GUI page information to LLM to elicit testing scripts, and executing them to keep passing the app feedback to LLM, iterating the whole process. 
Within it, we extract the static context of the GUI page and the dynamic context of the iterative testing process, design prompts for inputting this information to LLM, and develop a neural matching network to decode the LLM's output into actionable steps to execute the app.
We evaluate {\tool} on 86 apps from Google Play, and its activity coverage is 71\%, with 32\% higher than the best baseline, and can detect 36\% more bugs with faster speed than the best baseline. 
{\tool} also detects 48 new bugs on the Google Play with 25 of them being confirmed/fixed. 
We further summarize the capabilities of {\tool} behind the superior performance, including semantic text input, compound action, long meaningful test trace, and test case prioritization. 

\end{abstract}


\maketitle

\input{sec/introduction}
\input{sec/background}
\input{sec/approach}

\input{sec/effectiveness}
\input{sec/usefulness}
\input{sec/discussion}
\input{sec/related}
\input{sec/conclusion}
\input{sec/data.tex}
\bibliographystyle{ACM-Reference-Format}

\bibliography{reference}

\end{document}

%% file: macro.tex
\usepackage{soul}
\usepackage{float}
\usepackage{threeparttable}
\usepackage{booktabs}
\usepackage{multirow}
\usepackage{epstopdf}
\usepackage{hyperref}
\usepackage{listings}
\usepackage{fancybox}
\usepackage{graphicx}
\usepackage{subcaption}
\usepackage{paralist}
\usepackage{ragged2e}
\usepackage{enumitem}

\usepackage{marvosym}
\usepackage{multirow}
\usepackage{epstopdf}
\usepackage{hyperref}
\usepackage{listings}
\usepackage{fancybox}
\usepackage{graphicx}
\usepackage{subfloat}
\usepackage{cleveref}
\usepackage{paralist}
\usepackage{ragged2e}
\usepackage{enumitem}
\usepackage{pythonhighlight}
\usepackage{algpseudocode}

\usepackage[normalem]{ulem} 
\newcommand\hlg{\bgroup\markoverwith
  {\textcolor{green!10}{\rule[-.5ex]{2pt}{2.5ex}}}\ULon}
\newcommand\hlb{\bgroup\markoverwith
  {\textcolor{blue!10}{\rule[-.5ex]{2pt}{2.5ex}}}\ULon}
\newcommand\hlr{\bgroup\markoverwith
  {\textcolor{red!10}{\rule[-.5ex]{2pt}{2.5ex}}}\ULon}

\usepackage[framemethod=TikZ]{mdframed}
\usepackage{tcolorbox}
\makeatletter
\newcommand{\mybox}[1]{%
  \setbox0=\hbox{#1}%
  \setlength{\@tempdima}{\dimexpr\wd0+13pt}%
  \begin{tcolorbox}[boxrule=0.5pt, colback=white, arc=4pt,
      left=6pt,right=6pt,top=6pt,bottom=6pt,boxsep=0pt]
    #1
  \end{tcolorbox}
}
\usepackage{tikz}
\usepackage{pgfplots}
\usetikzlibrary{pgfplots.statistics,calc}
\usepackage{array}
\usepackage{amsmath}
\usepackage{centernot}
\usepackage{xspace}
\usepackage{url}
\usepackage{verbatim}
\usepackage{wrapfig}
\usepackage{tabularx}
\usepackage{bbding}
\usepackage{pifont}
\usepackage{wasysym}
\usepackage{amssymb}

\newcommand{\tool}{\texttt{GPTDroid}}

\newcommand{\chen}[1]{\textcolor{red}{#1}}

\makeatletter  
\newif\if@restonecol  
\makeatother  
\usepackage[linesnumbered,ruled,vlined]{algorithm2e}
\usepackage{algpseudocode}  
\usepackage{amsmath}  

%% file: sec/introduction.tex
\section{Introduction}
\label{sec_introduction}
Mobile apps have become increasingly popular over the past decade, with millions of apps available for download from app stores like the Apple App Store~\cite{Appstore} and Google Play Store~\cite{Googleplay}. 
With the rise of app importance in our daily life, it has become increasingly critical for app developers to ensure that their apps are of high quality and perform as expected for users. 
To avoid time-consuming and labour-extensive manual testing, automated GUI (Graphical User Interface) testing is widely used for quality assurance of mobile apps~\cite{mirzaei2016reducing,yang2018static,yang2013grey,machiry2013dynodroid,zeng2016automated,mao2016sapienz} i.e., dynamically exploring mobile apps by executing different actions such as scrolling, clicking based on the program analysis to verify the app functionality. 

However, existing GUI testing tools such as probability-based or model-based ones~\cite{su2021benchmarking,kong2018automated,rubinov2018we}  suffer from low testing coverage when testing practical commercial apps, meaning that they may miss important bugs and issues. 
This is because of the complex and dynamic nature of modern mobile apps~\cite{rubin2015covert,Rico,pecorelli2022software,rubinov2018we,kong2018automated,fan2018large}, which can have hundreds or even thousands of different screens, each with its own unique set of interactions and possible user actions and logic.
In addition, test inputs generated by these methods are significantly different from real users' interaction traces~\cite{peng2022mubot}, resulting in the low testing coverage. 
To address these limitations, there has been a growing interest in using deep learning (DL)~\cite{li2019humanoid,dong2020time,yang2022survey,yasin2021droidbotx} and reinforcement learning (RL)~\cite{pan2020reinforcement,romdhana2022deep,collins2021deep,lv2022fastbot2} methods for automated mobile GUI testing. 
By learning from human testers' behavior, these methods aim to generate human-like actions and interactions that can be used to test the app's GUI more thoroughly and effectively.
These approaches are based on the idea that the more closely the actions performed by the testing algorithm mimic those of a human user, the more comprehensive and effective the testing will be.

Nevertheless, there are still some limitations with these DL and RL based GUI testing methods.
First, learning algorithms require large amounts of data which is difficult to collect real-world users' interactions.
Second, learning algorithms are designed to learn and predict from training data, so they may not generalize well to the new, unseen situations, as apps are constantly evolving and updating.
Third, mobile apps can be non-deterministic, meaning that the outcome of an action may not be the same every time, it is performed (e.g., clicking the ``delete'' button from a list with the last content would produce an empty list for which the delete button no longer works)
which specifically makes it difficult for RL algorithms to learn and make accurate predictions.
Therefore, another more effective approach to generate human-like actions is highly needed to thoroughly test mobile apps.

The emerging Large Language Model (LLM)~\cite{attention,brown2020GPT3,chowdhery2022palm,zhang2022opt} 
trained on ultra-large-scale corpus, which shows promising performance in natural language understanding, logical reasoning and question answering in recent years. 
For example, GPT-3~\cite{brown2020GPT3} (Generative Pre-trained Transformer-3) is one LLM from OpenAI with 175 billion parameters trained on a massive dataset including existing test scripts and bug reports, which makes it capable of understanding and generating text across a wide range of topics and domains. 
The success of ChatGPT\footnote{https://openai.com/blog/chatgpt/} based on GPT-3 demonstrates that LLM can understand human knowledge and interact with humans as a knowledgeable expert.
Inspired by ChatGPT\footnote{ChatGPT cannot be directly applied for mobile GUI testing, so we adopt GPT-3 which provides official API and easy-to-control results.}
, we formulate GUI testing problem as a questions \& answering (Q\&A) task, i.e., asking the LLM to play a role as a human tester to test the target app.

In detail, we propose a new approach, {\tool}
, asking LLM to chat with mobile apps by passing the GUI page information to LLM to elicit testing scripts, and executing them to keep passing the app feedback to LLM, iterating the whole process.
To convert the visual information of the app GUI into the corresponding natural language description, we first extract the semantic information of the app and GUI page by decompiling the target app and view hierarchy files, as well as the dynamic context information of the iterative testing process.
 Similar to the screenreader through which the blind interact with mobile apps~\cite{Talkback, VoiceOver},
 we design linguistic patterns to generate prompts for describing the current GUI page as the input to LLM, which provides ways for LLM to interact with the app.
 Given the natural language described answer from LLM, we decode it into actionable steps to execute the target app by developing a neural matching network model.

\input{figure/example-of-prompt.tex}

Compared with conventional learning-based algorithms mentioned above, our approach is based on LLM as a zero-shot tester that does not require any training data or corresponding computational resources for training the model.
One example chat log can be seen in Figure~\ref{fig:Example-prompt}.
LLM can understand the app GUI, and provide detailed actions to navigate the app (e.g., A1-A5 at Figure~\ref{fig:Example-prompt}).
To compensate for its wrong prediction (A2 at Figure~\ref{fig:Example-prompt}), the real-time feedback by our approach guides it to regenerate the input until triggering a valid page transition.
It remains clear testing logic even after a long testing trace to make complex reasoning of actions (A3, A4 at Figure~\ref{fig:Example-prompt}), and it can prioritize to test important functions earlier (e.g., A5 at Figure~\ref{fig:Example-prompt}). 
A more detailed analysis of our approach's capability and reasons behind it is in Section~\ref{Sec_Discussion}.

To evaluate the effectiveness of {\tool}, we carry out an experiment on 86 popular Android apps in Google Play with 129 bugs.
Compared with 9 common-used and state-of-the-art baselines, {\tool} can achieve more than 32\% boost in activity coverage than the best baseline, resulting in 71\% activity coverage. 
As {\tool} can cover more activities, the method can detect 36\% more bugs with faster speed than the best baseline.
Apart from the accuracy of our {\tool}, we also evaluate the usefulness of our {\tool} by detecting unseen crash bugs in real-world apps from Google Play. 
Among 216 apps, we obtain 48 crash bugs with 25 of them being confirmed and fixed by developers, while the remaining are still pending. 
To reveal reasons behind the promising performance of our approach, we further investigate the experiment results qualitatively and summarize 4 findings in discussion including semantic text input, compound action, long meaningful test trace, and test case prioritization.

The contributions of this paper are as follows: 
\begin{itemize}

\item \textbf{Vision.} The first work to formulate the automated GUI testing problem to a 
question \& answering task by bringing LLM into GUI testing domain, as far as we know.

\item \textbf{Technique.} A novel approach {\tool}
based on ``pre-train, prompt and predict'' paradigm of the LLM by understanding the GUI semantic information and dynamic context of the iterative testing process, for automatically inferring possible operation steps.

\item \textbf{Evaluation.} Effectiveness and usefulness evaluation of {\tool} in the real-world apps with practical bugs detected.

\item \textbf{Insight.} Detailed qualitative analysis in discussion revealing the reasons why LLM can generate human-like actions for app testing. 

\end{itemize}



%% file: figure/example-of-prompt.tex
\begin{figure}[t]
\centering
\vspace{0.05in}
\includegraphics[width=8.5cm]{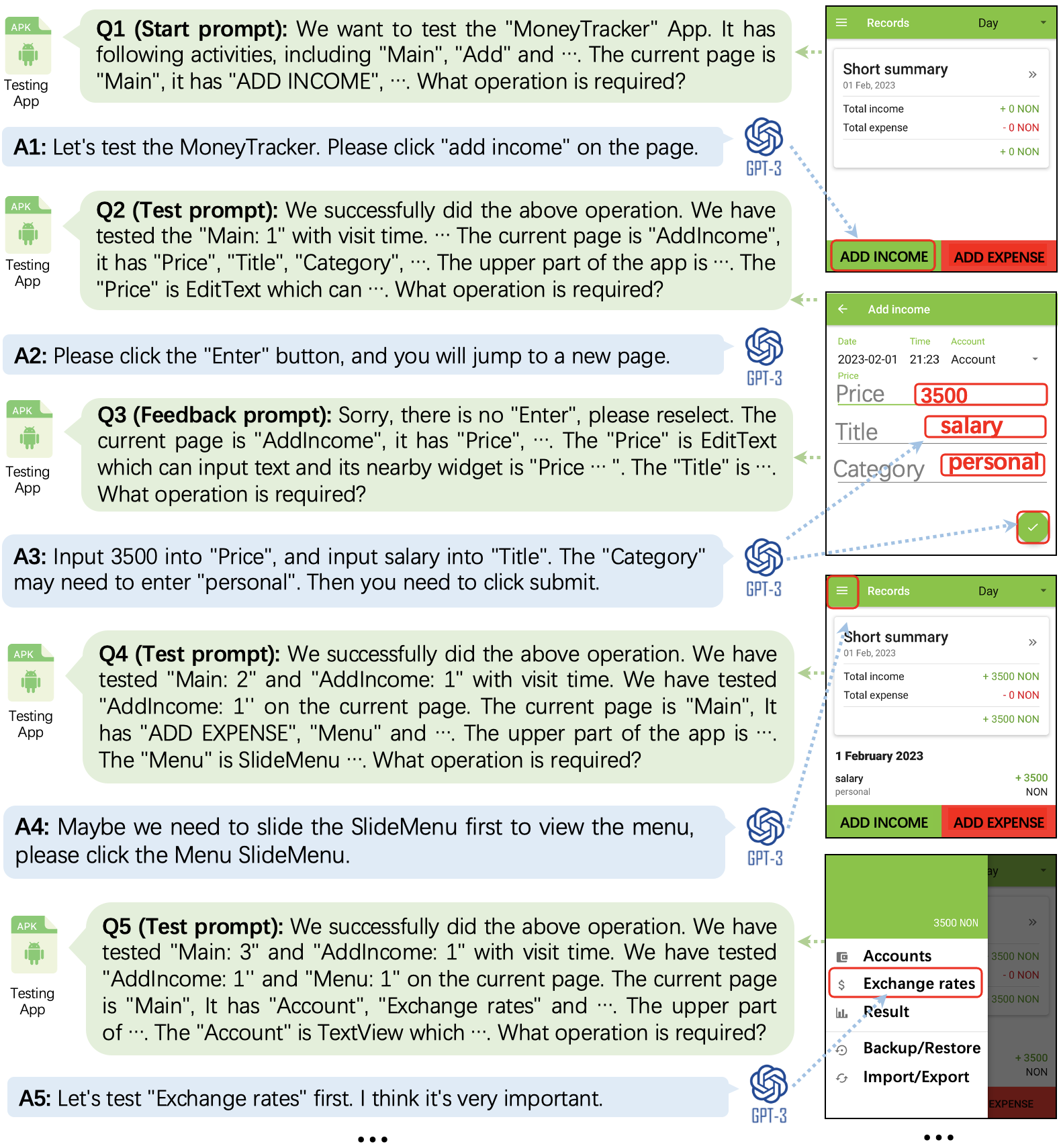}
\caption{Demonstrated example of how {\tool} works.}
\label{fig:Example-prompt}
\vspace{-0.1in}
\end{figure}

%% file: sec/background.tex
\section{Background}
\label{sec_background}
\subsection{Android GUI and GUI event}
\label{sec_background_UI_GUI_event}
For a mobile app, the user interface (UI) is the place where interactions between humans and machines occur. App developers design UI to help users understand the features of their apps, and users interact with the apps through the UI. 
To help developers manipulate views flexibly, Android Software Development Kit (SDK)~\cite{Android} allows developers to build UI in Android source code using the View and ViewGroup objects. 
The View objects are usually called ``widget'' (e.g., \textit{ImageView}, \textit{TextView}), while the ViewGroup objects are usually called ``layout'' which provides various layout structures (e.g., \textit{LinearLayout}, \textit{RelativeLayout}).
Every widget and layout in the Android source code has its specific attributes, which are used to set its boundaries, clickable and referenced external resources, etc.
During the running process, developers can obtain the view hierarchy file corresponding to the current UI page (screenshot) through ``uiautomator dump'' of the Android Debug Bridge (ADB) command~\cite{Android}. The view hierarchy file includes the widget information (coordinate information, ID, widget type, text description, etc.), and the layout information on the current UI page~\cite{Viewhierachy}, which can be used by automated GUI testing tools to obtain the information about widgets.

The graphical user interface (GUI) is the most important type of UI for most mobile apps, where apps present content and actionable widgets on the screen and users interact with the widgets using actions (GUI event) such as clicks, swipes, and text inputs. In the process of using apps, users often need complex GUI events to jump from one page to another, such as filling in multiple text input widgets, sliding widgets to the left to operate, and long pressing widgets to delete.

\subsection{Large Language Model \& Prompt }
\label{sec_motivation_Large_Language_Model}
Pre-trained Large Language Models (LLMs) have been shown effective in many natural language processing (NLP) tasks. It is trained on ultra-large-scale corpus and can understand the input prompts (sentences with prepending instructions or a few examples) and generate reasonable text.
When pre-trained on billions of samples from the Internet, recent LLMs (like GPT-3~\cite{brown2020GPT3}, PaLM~\cite{chowdhery2022palm} and OPT~\cite{zhang2022opt}) encode enough information to support many NLP tasks~\cite{lucy2021gender,sharples2022automated, yang2022empirical}.
GPT-3~\cite{brown2020GPT3} is one of the most popular and powerful LLM which has great performance in many text generation tasks. 
It is based on the transformer model~\cite{attention} including input embedding layers, masked multi-self attention, normalizaiton layers, and feed-forward in Fig \ref{fig:GPT-3}. 
Given a sentence, the input embedding layer encodes it through the word embedding. The multi-self attention layer is used to divide a whole high-dimensional space into several different subspaces to calculate the similarity. The normalization layer is implemented through a normalization step that fixes the mean and variance of each layer's inputs. The feed-forward layer compiles the data extracted by previous layers to form the final output. 

\input{figure/GPT-3.tex}

Recently, prompt engineering has been proposed to close the gap between pre-training and downstream tasks. Instead of designing a new training objective for each downstream task, prompt engineering rewrites the input by adding a natural language instruction such as ``This is XX app, On its xxx page, it has xxx. What to do next?'' to reuse the masking objective for downstream tasks. 
Formally, a popular prompt engineering employs a prompt template $T_{prompt}(.)$ to convert the input $X$ to prompt input $X_{prompt} = T_{prompt}(X)$. The prompt template is a textual string with unfilled slots to fill the input $X$ and a question.

For automated GUI testing test script generation, the filled input $X$ is the GUI information of the page, and LLM attempts to generate the operation steps to be executed. 
Although achieving promising results in various NLP tasks, the standard prompt is ineffective for automated GUI testing, because LLM cannot understand the visual information of the GUI page or the source code of the app. 
Therefore, this paper focuses on how to describe the GUI page information to let LLM better understand, and how to decode the LLM's feedback to actionable operation steps to execute the app. 


%% file: figure/GPT-3.tex
\begin{figure}[htb]
\centering
\vspace{-0.05in}
\includegraphics[width=8.3cm]{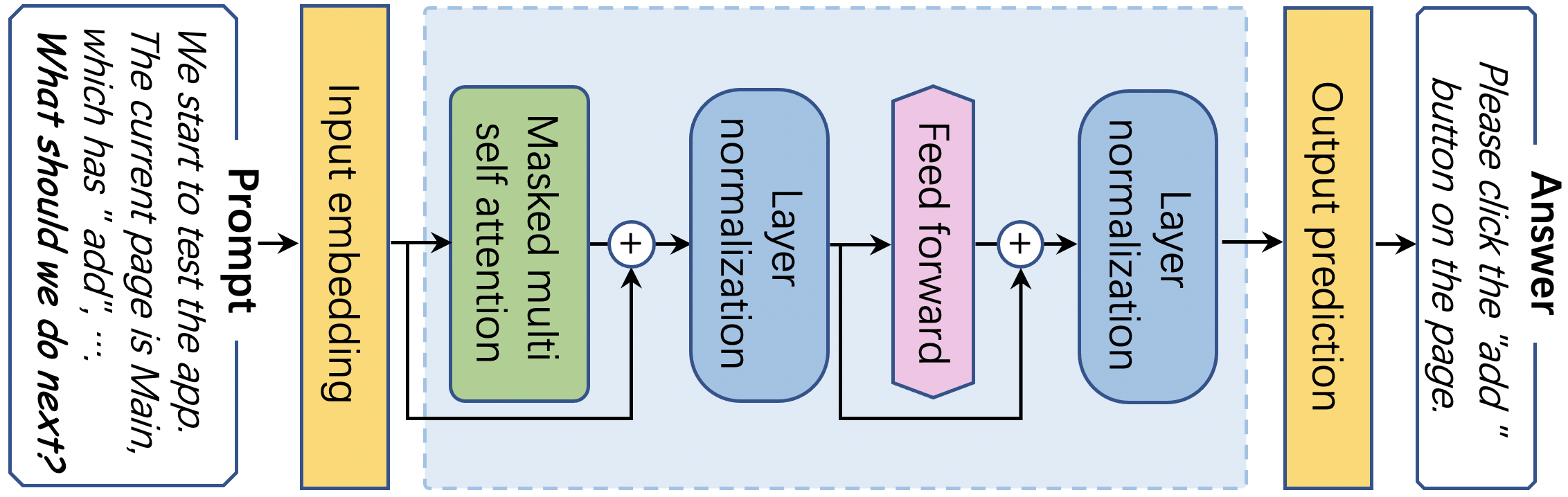}
\vspace{-0.05in}
\caption{The model structure of GPT-3.}
\label{fig:GPT-3}
\vspace{-0.05in}
\end{figure}

%% file: sec/approach.tex
\section{Approach}
\label{sec_approach}

We model the GUI testing as a Question \& Answering (Q\&A) problem, i.e., asking the LLM to play a role as a human tester, and enabling the interactions between the LLM and the app under testing.
To realize this, we propose {\tool}, as demonstrated in Figure \ref{fig:overview}.
It extracts the static and dynamic context of the current GUI page, encodes them into prompt questions for LLM, decodes LLM's feedback answer into actionable operation scripts to execute the app; and iterates the whole process. 
Armed with the knowledge learned from large-scale training corpus, {\tool} would have the potential to guide the testing in exploring more diversified pages, conducting more complex operational actions, and covering more meaningful operational sequences.  

\input{figure/overview.tex}

\input{tab/App-GUI-information.tex}
Specifically, in each iteration of the testing, {\tool} first obtains the view hierarchy file of the mobile app and extracts the static context including the app information, information of the current GUI page, and details of each widget in the page. 
It also maintains a testing operation memorizer, and extracts the latest dynamic information from it to indicate current testing progress.
Based on this contextual information, we design linguistic patterns for generating the prompts as input of LLM, and the LLM would output the operational steps described in natural language. 
We then design a natural matching network to match the operational steps with the GUI events (i.e., widgets) of the app to enable it to be automatically executed.


\subsection{Context Extraction}
\label{subsec_approach_information_Extraction}

Despite of its excellence on various tasks, the performance of LLM can be significantly influenced by the quality of its input, i.e., whether the input can precisely describe what to ask~\cite{chen2022knowprompt,liao2022ptau,zhou2022learning}.
In the scenario of this interactive mobile GUI testing, we need to accurately depict the GUI page currently under test, as well as its contained widgets information from a more micro perspective, and the app information from a more macro perspective.
Furthermore, to act like a human tester, {\tool} should also capture current testing progress so as to recommend the testing operations from a more global viewpoint to potentially cover more activities and avoid duplicate explorations. 
This section describes which information will be extracted, organized into static context and dynamic context to facilitate reading. 
And Section \ref{subsec_approach_Prompt_Generation} will describe how we organize this information into the style that LLM can better understand.


\subsubsection{\textbf{Static Context Extraction}}
\label{subsubsec_approach_Information_Extraction}

Static context relates to the information of the app, the GUI page currently tested, and all the widgets on the page. 
The app information is extracted from the \textit{AndroidMaincast.xml} file, while the other two types of information are extracted from the view hierarchy file, which is obtainable with UIAutomator~\cite{uiautomator}.
Table \ref{tab:app-GUI-info} presents the summarized view of them. 

\textbf{App information} provides the macro-level semantics of the app under testing, which facilitates the LLM to gain a general perspective about the functions of the app. 
The extracted information includes the name of the app and the name of all its activities.

\textbf{Page GUI information} provides the semantics of the current page under testing during the interactive process, which facilitates the LLM to capture the current snapshot.  
We extract the activity name of the page,  all the widgets represented by the ``text'' field or ``resource-id'' field (the first non-empty one in order), and the widget position of the page.
For the position, inspired by the screen reader~\cite{zhang2021screen,stangl2021going,wang2021screen2words}, we first obtain the coordinates of each widget in order from top to bottom and from left to right, and the widgets whose ordinate is below the middle of the page is marked as lower, and the rest is marked as upper.

\textbf{Widget information} denotes the micro-level semantics of the GUI page, i.e., the inherent meaning of all its widgets, which facilitates the LLM in providing actionable operational steps related to these widgets. 
The extracted information includes ``text'', ``hint-text'', and ``resource-id'' field (the first non-empty one in order), ``class'' field, and ``clickable'' field.
We also extract the information from nearby widgets to provide a more thorough perspective, which includes the ``text'' of parent node widgets and sibling node widgets.

\input{tab/approach-rule.tex}

\subsubsection{\textbf{Dynamic Context Extraction}}
\label{subsubsec_approach_Information_Extraction}

Dynamic context relates to the detailed testing progress, which facilitates the LLM being well aware of the process context and making informed decisions.  
We design an \textbf{operation memorizer} to keep the record of this information, i.e., whether and how many times a GUI page has been explored or a widget has been operated, as shown in Table \ref{tab:app-GUI-info}.

Specifically, during the iteration, when an operation is conducted, we can obtain the widget information of the operation, and the GUI pages information after the operation, and then the operation memorizer is updated accordingly. 
In detail, the visit number of the widget is updated through finding the same widget in the operation memorizer by the ``text'' field and ``resource-id'' field of the widget.
The visit number of the GUI page is updated through finding the same activity in the memorizer with the ``ActivityName'' field of the page.

\subsection{\textbf{Prompt Generation}}
\label{subsec_approach_Prompt_Generation}

With the extracted information, we design linguistic patterns to generate prompts for inputting into the LLM. 

We first conduct preprocessing for the extracted information, to facilitate the follow-up design. 
For each static attribute in Table \ref{tab:app-GUI-info}, we tokenize it by the underscore and Camel Case~\cite{pascalcase} (e.g. Capital letters of each word) considering the naming convention in app development and remove the stop words to reduce noise.
We then conduct the part-of-speech (POS) tagging with Standford NLP parser~\cite{de2008stanford}, and only retain the noun, verb and prepositions for the linguistic patterns.

\subsubsection{\textbf{Linguistic Patterns of Prompt}}
\label{subsubsection_patterns_of_prompt}

To design the patterns, each of the five authors is asked to write the prompt sentence following regular prompt template~\cite{chen2022knowprompt,Cantino201Prompt,gu2021ppt}, and questions the LLM for generating the operation steps. 
He/she then checks to what extent the recommended operation is reasonable considering the whole testing process.
Each author can access a random-chosen 100 apps from Google play, and he/she can obtain the preprocessed static context information and the dynamic context information. 
After 10 hours of trial, he/she is required to provide the most promising and diversified 20 prompt sentences, which are served as the seeds for designing patterns. 
With the prompt sentences, the five authors then conduct card sorting~\cite{spencer2009card} and discussion to derive the linguistic patterns.
As shown in Table \ref{tab:approach-rule}, this process comes out with 6 linguistic patterns corresponding with the four sub-types of information in Table \ref{tab:app-GUI-info} and two operation \& feedback patterns.

\textbf{Pattern related to static context:} We design three patterns to describe the overview of the GUI page currently under testing, respectively corresponding to the app information, page GUI information, and widget information in Table \ref{tab:app-GUI-info}.

\textbf{Pattern related to dynamic context:} We design one pattern to describe the testing progress with the dynamic context as shown in Table \ref{tab:app-GUI-info}.

\textbf{Pattern related to operation \& feedback question:} We design two patterns to describe the operational and feedback question. For the operational question, we ask the LLM what operation is required. And for the feedback question, after deciding the previous operation is not applicable (as described in Section \ref{subsec_approach_Output-Matching}), we inform the LLM that there is no such widget on the current page, and let it re-try.

\subsubsection{\textbf{Prompt Generation Rules:}}
\label{subsubsection_Prompt_generation_rules}

Since the designed patterns describe information from different points of view, we combine the patterns from different viewpoints and generate the prompt rules as shown in Table \ref{tab:approach-rule}.
We design three kinds of prompts respectively for starting the test, routine inquiry, and the feedback in case of error occurred.
Note that, due to the robustness of the LLM, the generated prompt sentence does not need to follow the grammar completely.

\textbf{Test prompt} is the most commonly used prompt for informing the LLM of the current status and query for the next operation. 
Specifically, we first tell the LLM the dynamic context, i.e., how many times each GUI page and widget has been explored; followed by the static context, i.e., the information about the current GUI page and detailed widget information; then ask the LLM which operation is required.

\textbf{Feedback prompt} is used for informing the LLM error occurred and re-try for querying the next operation.
Specifically, we first tell LLM its generation operation cannot correspond to the widget on the page; re-provide it the detailed widget information of the page and let the LLM recommend the operation again.

Besides the above two kinds of prompts, we additionally design \textbf{start prompt} to start the testing of the app. 
Different from the test prompt, it provides the LLM with the information about the app with all activities for a global overview; and it does not have dynamic context since the testing just begins.
This prompt is only used once since the LLM can somehow remember this global app information during the testing process~\cite{su2017guided,li2019humanoid}.

\subsection{Operation Matching}
\label{subsec_approach_Output-Matching}

After inputting the generated prompt, LLM would output a natural language sentence describing the operation steps for the testing, e.g., click the save button.
We need to convert the natural language described operation steps to the GUI events (i.e., widgets) of the app to enable it to be automatically executed. 
This is non-trivial considering the natural language description can be arbitrary, and inherently imprecise. 
We design a neural matching network to predict which widget can be most likely to be mapped to the operation step. 
Since training the neural network usually requires a large amount of labeled data, we develop a heuristic-based automated training data generation method to facilitate the model training in Section \ref{subsubsec-Heuristic-based Training Data Generation}.

\input{figure/matching-network.tex}

\subsubsection{\textbf{Neural Matching Network}}

As shown in Figure~\ref{fig:matching-network}, 
one input of the neural matching network is the LLM's feedback answer, which is the natural language described operation step $C_{step}$, while the other input is the textual information $C_{text}$ of app's widget.
We choose the first non-empty ``text'', ``ID'', and ``description'' fields of the widget in turn. 
The operation step $C_{step}$ and the textual information of the widget $C_{text}$ are concatenated with symbol \textit{$\textless$SEP$\textgreater$}, then input into the pre-trained transformer encoder to generate the hidden state of the text.
Finally, the textual hidden state is input into a fully connected layer for obtaining the matching score of the feedback answer and the widget. 

During the iterative testing process, each time when the LLM provides the feedback answer, {\tool} would first separate the answer with \textit{``and'', ``,'', ``.''} to derive the atomic operation step, considering the compound operations can be recommended by the LLM.
Then for each atomic operation step, we compute the matching score with all candidate widgets in the current GUI page, and choose the widget with the highest matching score as the target widget. We then use the \textit{WidgetAction} attribute of the widget in Table \ref{tab:app-GUI-info} to performance the action on the target widget for executing the app.
Besides, if the matching score of all widgets is less than 0.5, we determine there is no satisfied widget on the page. 
It indicates an error occurred in the LLM's feedback answer, and would activate the feedback prompt in the next iteration.

\subsubsection{\textbf{Heuristic-based Training Data Generation}}
\label{subsubsec-Heuristic-based Training Data Generation}
Training such a neural matching network requires a large amount of labeled data with natural language described operation step and GUI event, e.g., \textit{press the back button} and corresponding event \textit{click back button} in the app.
However, there is no such open dataset, and collecting it from scratch is time- and effort-consuming. 
Meanwhile, by examining the operation step of LLM's output, we observe that they tend to follow certain linguistic patterns. 
This motivates us in developing a heuristic-based automated training data generation method for collecting the satisfied training data.

The primary idea is that for each interactive widget in a GUI page, we heuristically generate the operation step which can operate on it for transitioning to the next state;
meanwhile, generate the negative data instances between the operation step and the irrelevant widgets.
To derive the heuristic rules, the five authors examine 400 operation steps of LLM's output, summarizing the linguistic patterns for writing the operation steps.
We have summarized 31 non-repetitive operation descriptions and their variants, such as click, enter, press, etc. 
We uploaded the complete table to our website\textsuperscript{\ref{github}}.
For each iterative widget, we randomly generate three different operation descriptions to serve as the positive data instances in the training data. 
For the negative data instances, we follow the hard negative mining strategy~\cite{liu2016ssd} to enhance the discriminability of the model.

\subsection{Implementation}
\label{subsec_approach_Implement}

Our {\tool} is implemented as a fully automated GUI app testing tool, which uses or extends the following tools: VirtualBox~\cite{virtualbox} and the Python library pyvbox~\cite{pyvbox} for running and controlling the Android-x86 OS, Android UIAutomator~\cite{uiautomator} for extracting the view hierarchy file, and Android Debug Bridge (ADB)~\cite{Adb} for interacting with the app under test (Section \ref{subsec_approach_information_Extraction}).

For the LLM (Section \ref{subsec_approach_Prompt_Generation}), we use 
the pre-trained GPT-3 model which was released on the OpenAI website\footnote{\url{https://beta.openai.com/docs/models/gpt-3}}.  
The basic model of GPT-3 is the \textit{text-davinci-003} model which is extremely powerful and good at answering questions.
We build our neural matching network model (Section \ref{subsec_approach_Output-Matching}) based on PyTorch~\cite{pytorch} and Sentence Transformers~\cite{sbert}. 
The text processing module is loaded with DistilBert~\cite{sanh2019distilbert}, a 12-layer transformer-based pre-training model. We use AdamW as the optimizer, BCEWithLogitsLoss as the loss function, and train the model with the batch size set to 20. 

%% file: figure/overview.tex
\begin{figure}[!th]
\centering
\includegraphics[width=8.6cm]{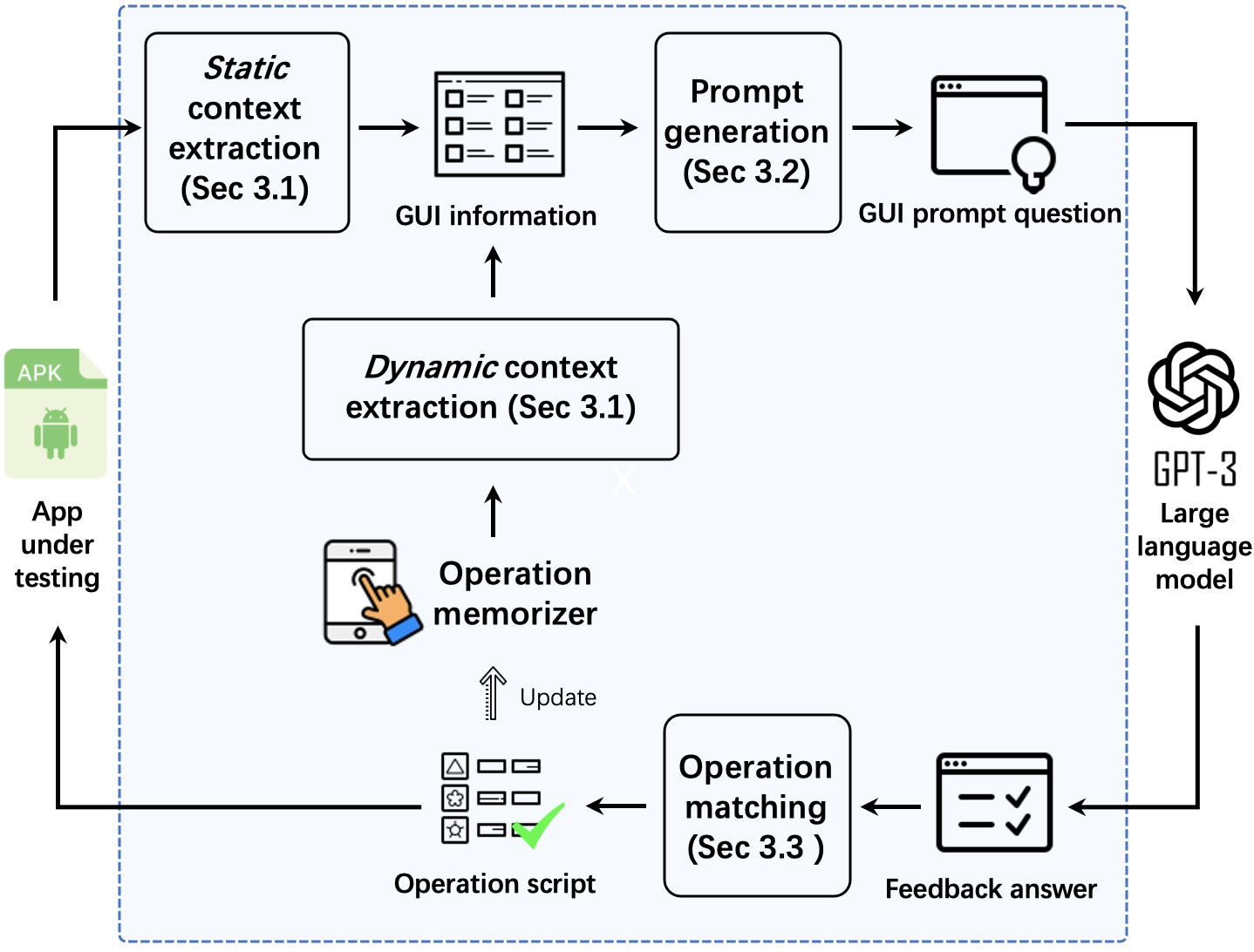}
\caption{Overview of {\tool}.}
\label{fig:overview}
\vspace{-0.15in}
\end{figure}

%% file: tab/App-GUI-information.tex
\begin{table*}
\vspace{0.15in}
\caption{Extracted information and examples.}
\vspace{-0.1in}
\label{tab:app-GUI-info}
\centering
\footnotesize
\begin{center}
\begin{tabular}{m{0.3cm}<{\centering} | m{1.6cm}<{} | m{8.4cm}<{} | m{6.0cm}<{}}
\toprule
\textbf{Id} & \textbf{Attribute} & \textbf{Description}  & \textbf{Examples} \\
\midrule
\multicolumn{4}{c}{\textbf{Static context - App information}} \\
\midrule
1 & AppName & Name of the app under testing & AppName = ``Money Tracker'' \\
2 & Activities & List of names for all activities of the app, obtained from \textit{AndroidManifest.xml} file & Activities = [``Main'', ``AddAccount'', ``Import'', ``Income'', ...] \\
\midrule
\multicolumn{4}{c}{\textbf{Static context - page GUI information}}\\ 
\midrule
3 & ActivityName & Activity name of the current GUI page & ActivityName = ``AddPersonalInformation'' \\
4 & Widgets & List of all widgets in current page, represented with text/id & Widgets = [``Edit Account'', ``btn\_income'', ...] \\
5 & Position & Relative position of widgets, obtained through their coordinates & Upper = [``Welcome'', ...], Lower = [``Add Income'', ...] \\
\midrule
\multicolumn{4}{c}{\textbf{Static context - widget information}} \\ 
\midrule
6 & WidgetText & Widget text, obtained by field `text' or `hint-text' & WidgetText = ``Welcome to the Money Tracker!'' \\
7 & WidgetID & Widget ID, obtained by field `resource-id'. & WidgetID = ``add\_account'' \\
8 & WidgetCategory & Category: TextView, EditText, ImageView, etc, obtained by field `class' & WidgetCategory = ``TextView'' \\
9 & WidgetAction & Widget action, obtained by field `clickable', such as click, input, etc. & WidgetAction = ``Click'' \\
10 & NearbyWidget & Nearby widgets, obtained by the text of parent widgets and sibling widgets & NearbyWidget = ``your income: [SEP] \$ '' \\
\midrule
\multicolumn{4}{c}{\textbf{Dynamic context}} \\ 
\midrule
11 & PageVisits & Set of tested GUI pages with page visits number & PageVisits = [\{``Main'': ``5'' , ``Account'': ``3'', ``Setting'': ``1'',  ...\}]\\
12 & WidgetVisits & Set of tested widgets of current GUI page with widget visits number & WidgetVisits = [\{``Income'': ``2'', ``Add'': ``2'', ``Delete'': ``3'', ...\}]\\

\bottomrule
\end{tabular}
\end{center}
\vspace{-0.05in}
\end{table*}

%% file: tab/approach-rule.tex
\begin{table*}
\vspace{0.1in}
\caption{The example of linguistic patterns of prompts and prompt generation rules.}
\vspace{-0.1in}
\label{tab:approach-rule}
\centering
\footnotesize
\begin{center}
\begin{tabular}{m{0.3cm}<{\centering} | m{8.1cm}<{} | m{7.9cm}<{}}
\toprule
\textbf{Id} & \textbf{Sample of linguistic patterns/rules} & \textbf{Instantiation}\\
\midrule
\multicolumn{3}{c}{\textbf{Static context patterns: \hlr{$\langle StaticContext\rangle$}}}\\ 
\midrule
\rowcolor{red!10}
1 & We want to test the \textit{`AppName'} App. It has the following activities, including \textit{`Activities'}. & We want to test ``Money tracker'' App. It has the following activities, including ``Main'', ``AddAccount'', ``Import'', ``Setting'', ... . \\ 
\rowcolor{red!10}
2 & The current page is \textit{`ActivityName'}, it has \textit{`Widgets'}. The upper part of the app is \textit{`Position'}, the lower part is \textit{`Position'}. & The current page is ``Main'', it has ``Income'', ``Add'', ... . The upper part of the app is ``Welcome to ..., delete, ...'', the lower part of the app is ``Income, add, ...''.  \\ 
\rowcolor{red!10}
3 & The widgets which can operation are \textit{`WidgetText / WidgetID'}. \textit{`WidgetText / WidgetID'} is \textit{`WidgetCategory'} which can \textit{`WidgetAction'} and its nearby widget is \textit{`NearbyWidget'}.  & The widgets which can operation are ``add'', ``delete'', ... . ``add'' is Button which can click and its nearby widget is  ``Add account, ...'' , ``delete'' is TextView which can click and its nearby widget is ... . \\ 
\midrule
\multicolumn{3}{c}{\textbf{Dynamic context patterns: \hlb{$\langle DynamicContext\rangle$}}}\\ 
\rowcolor{blue!10}
\midrule
4 & We have tested \textit{`PageVisits'} with visit time. We have tested the \textit{`WidgetVisits'} on the current page. 
& We have tested ``Main: 7'', ``About: 2'', ``Account: 5'', ... with visit time. We have tested the ``Add: 1'', ``Delete: 2'', ``Edit the order: 1'' on the current page.\\
\midrule
\multicolumn{3}{c}{\textbf{Operation \& feedback question patterns: \hlg{$\langle OperationQuestion\rangle $}}} \\  
\midrule
 \rowcolor{green!10}
5 & What operation is required?  & What operation is required?  \\ 
 \rowcolor{green!10}
6 & There is no \textit{`WidgetText / WidgetID'} on the current page, please reselect.  & There is no ``Input'' on the current page, please reselect. \\ 
\bottomrule
\toprule
\multicolumn{3}{c}{\textbf{Prompt generation rules}}\\
\midrule
1 & \textbf{Start Prompt:} \hlr{$\langle StaticContext\rangle$[1,2,3]} + \hlg{$\langle OperationQuestion\rangle $[5]} 
& \hlr{We want to test “Money tracker” App, It has the following activities, including ``Main'', ``AddAccount'', ``Exchange'', ... . The current page is ``Main'', it has ``Income'', ``Add'', ... . The upper of the app is ``Welcome to ..., delete, ...'', the lower of the app is ``Income, add, ...''. The ``Income'' is Button which can click and its nearby widget is "Please input ...", ... .} \hlg{What operation is required?} \\
\midrule
2 & \textbf{Test Prompt:} We successfully did the above operation. \hlb{$\langle DynamicContext\rangle$[4]} + \hlr{$\langle StaticContext\rangle$[2,3]} + \hlg{$\langle OperationQuestion\rangle $[5]} 
& We successfully did the above operation. \hlb{We have tested the ``Main: 4'' and ``Exchange: 2'' with visit time. We have tested ``Add Exchange: 2'' on the current page.} \hlr{The current page is ``AddAccount'', it has ``Exchange'', ``Edit'', ... . The upper part of the app is ``delete, Income info ...'', the lower part of the app is ... The ``Exchange'' is Button which can click and its nearby widget is "Exchange the ....".} \hlg{What operation is required?} \\
\midrule
3 & \textbf{Feedback Prompt:} Sorry,  \hlg{$\langle OperationQuestion\rangle$[6]} + \hlr{$\langle StaticContext\rangle$[3]} + \hlg{$\langle OperationQuestion\rangle $[5]} 
& Sorry, \hlg{there is no ``Enter'' button on this page, please reselect.} \hlr{The current page is ``AddAccount'', it has ``Exchange'', ``Edit'', ... . The ``Exchange'' is TextView which can click and its nearby is "Exchange the ....".} \hlg{What operation is required?} \\
\bottomrule
\end{tabular}
\begin{tablenotes}
\footnotesize
\item \textbf{\textit{Notes:}} ``[1,2, ..., 5]'' means the id of each pattern. 
\end{tablenotes}
\end{center}
\vspace{-0.1in}
\end{table*}

%% file: figure/matching-network.tex
\begin{figure}[htb]
\centering
\vspace{-0.05in}
\includegraphics[width=8.2cm]{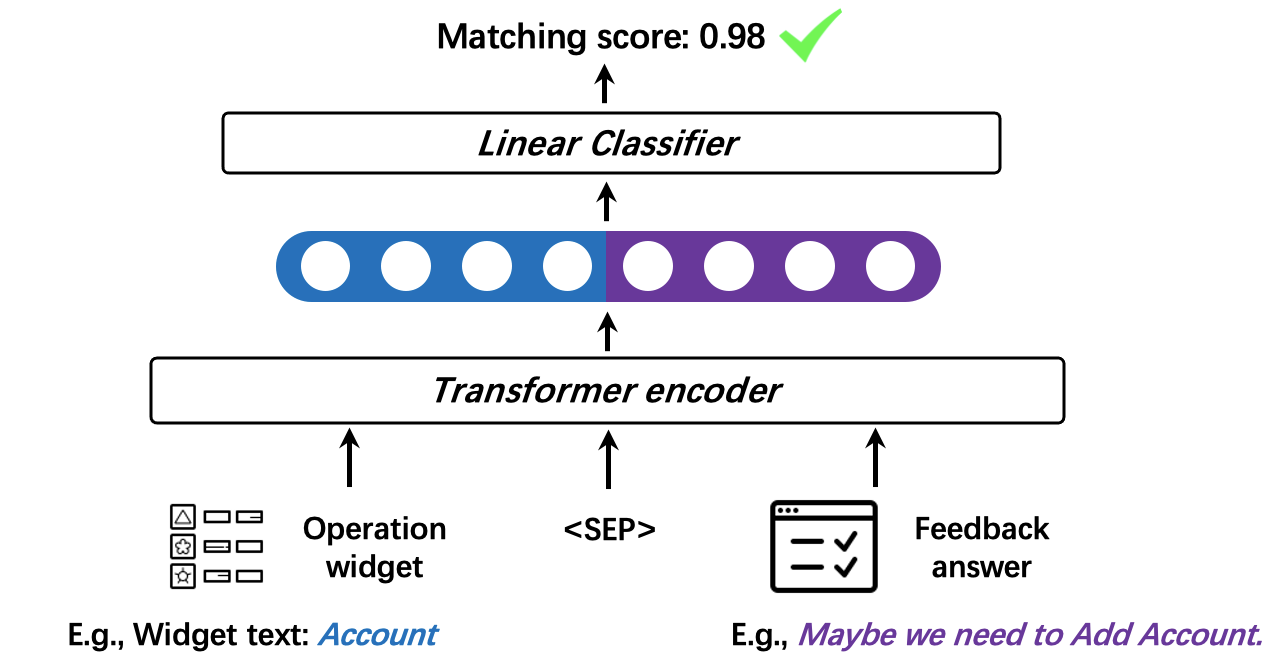}
\vspace{-0.05in}
\caption{The matching network architecture.}
\label{fig:matching-network}
\vspace{-0.1in}
\end{figure}

%% file: sec/effectiveness.tex
\section{Effectiveness Evaluation}
\label{sec_Effectiveness}
In order to verify the performance of {\tool}, we evaluate it by investigating the activity coverage (RQ1) and the number of detected bugs (RQ2).
In addition, we present the performance of operation matching in RQ3. 
Note that, this section utilizes the previously-detected bugs in the app's repositories to demonstrate the effectiveness of our approach, and the next section will further evaluate the usefulness of {\tool} in detecting new bugs. 







\subsection{Experimental Setup}
\label{subsec_experiment_dataset}
The experimental dataset comes from two sources. 
The first is from the apps in Themis benchmark ~\cite{su2021benchmarking}, which contains 20 open-source apps with 34 bugs in GitHub. 
Considering the small number of apps in the benchmark, we collect a second dataset following similar procedures as the benchmark.

In detail, we crawl the 50 most popular apps of each category from Google Play~\cite{Googleplay}, and we keep the ones with at least one update after May. 2022, resulting in 317 apps in 12 Google Play categories. 
Then, we use 9 common-used and state-of-the-art automated GUI testing tools (details are in Section \ref{subsec_experiment_baseline}) to run these apps in turn to ensure that they work properly. We then filter out the unusable apps by the following criteria: (1) UIAutomator~\cite{uiautomator} can't obtain the view hierarchy file; (2) they would constantly crash on the emulator; (3) one or more tools can't run on them; 
(4) The registration and login functions cannot be skipped with scripts~\cite{su2021benchmarking,lv2022fastbot2,dong2020time}; 
(5) They don't have issue records or pull requests on GitHub. 

There are 66 apps (with 93 bugs) remaining for this effectiveness evaluation.
Note that, same as the benchmark, all bugs are crash bugs.
Specifically, for each app, we select the version in which the bugs are confirmed by developers (merged GitHub pull requests) as our experimental data, following the practice of the benchmark.
The details of all 86 experimental apps (20 + 66) and related bugs are shown in Table \ref{tab:app-info}. 

\input{tab/dataset.tex}

Note that, there are 101 apps that are filtered out for effectiveness evaluation, yet can successfully run with our proposed approach. 
We apply them to the manual prompt generation in Section \ref{subsubsection_patterns_of_prompt} and heuristic training data generation in Section~\ref{subsubsec-Heuristic-based Training Data Generation}.
And this ensures there is no overlapping between the apps in approach design and evaluation.


We employ activity coverage and the number of detected bugs, which are widely used metrics for evaluating GUI testing~\cite{he2020textexerciser,liu2017automatic,arnatovich2018mobolic}.
We also present the number of covered activities and widgets which are also commonly-used metrics~\cite{li2019humanoid,pan2020reinforcement,su2021benchmarking} in Table \ref{tab:RQ1-result}.


\subsection{Baselines}
\label{subsec_experiment_baseline}
To demonstrate the advantage of {\tool}, we compare it with 9 common-used and state-of-the-art baselines. 
We roughly divide the GUI testing approaches into random-/rule-based methods, model-based methods, and learning-based methods, to facilitate understanding.

For random-/rule-based methods, we use Monkey~\cite{Monkey} and Droidbot~\cite{li2017droidbot}. For model-based methods, we use Stoat~\cite{su2017guided}, Ape~\cite{gu2019practical}, Fastbot~\cite{cai2020fastbot}, ComboDroid~\cite{wang2020combodroid}, TimeMachine~\cite{dong2020time}. For learning-based methods, we use Humanoid~\cite{li2019humanoid} and Q-testing~\cite{pan2020reinforcement}. 

We deploy the baselines and our approach on a 64-bit Ubuntu 18.04 machine (64 cores, AMD 2990WX CPU, and 128GB RAM) and evaluate them on Google Android 7.1 emulators (API level 25). Each emulator is configured with 2GB RAM, 1GB SDCard, 1GB internal storage, and X86 ABI image. Different types of external files (including PNGs / MP3s / PDFs / TXTs / DOCXs) are stored on the SDCard to facilitate file access from apps. 
Following common practice~\cite{gu2019practical,li2017droidbot}, we registered separate accounts for each bug that requires login and wrote the login scripts, and during testing reset the account data before each run to avoid possible interference. In order to ensure fair and reasonable use of resources, we set up the running time of each tool in one app to 30 minutes, which is widely used in other GUI testing studies~\cite{fan2018large,su2021benchmarking,gu2019practical,li2017droidbot}.
We run each tool three times and obtain the highest performance to mitigate potential bias.

\input{tab/RQ1-result.tex}

\subsection{Results and Analysis}
\label{subsec_results}
\subsubsection{\textbf{Performance of Activity Coverage (RQ1)}}
\label{sec_results_RQ1}

Table \ref{tab:RQ1-result} shows the number of covered widgets, number of covered activities, and average activity coverage of {\tool} and the baselines. 
We can see that {\tool} covers far more widgets and activities than the baselines, and the average activity coverage achieves 71\% across the 86 apps. 
It is 32\% (0.71 vs. 0.54) higher even compared with the best baseline (TimeMachine).
This indicates the effectiveness of {\tool} in covering more activities and widgets, thus bring higher confidence to the app quality and potentially uncovering more bugs. 

\input{figure/RQ1-ActivityCoverage.tex}

Figure \ref{fig:RQ1-ActivityCoverage} additionally demonstrates the average activity coverage with varying time. 
We can see that, in every time point, {\tool} achieves higher activity coverage than the baselines, and it achieves high coverage within about 13 minutes.
This again indicates the effectiveness and efficiency of {\tool} in covering more activities with less time, which is valuable considering the testing budget.

Among the baselines, the model-based and learning-based approaches have relatively higher performance. 
Yet the model-based approaches cannot capture the GUI semantic information and the exploration could not well understand the inherent business logic of the app.
In addition, existing learning-based approaches only use few context information for guiding the exploration, and the learners only have limited intelligence restricted by the model architecture and amount of labeled training data.

We further analyze the potential reasons for the uncovered cases.
First, some widgets or inputs do not have meaningful ``text'' or ``resource-id'', which hinders the approach of effectively understanding the GUI page. 
Second, some app requires specific operations, e.g., database connection, long press and drag widgets to a fixed location, which is difficult if not impossible to be automatically achieved.

\subsubsection{\textbf{Performance of Bug Detection (RQ2)}}
\label{sec_results_RQ2}


Figure \ref{fig:RQ2-result} shows the overall number of detected bugs of {\tool} and baselines with varying times. 
{\tool} detects 72 bugs for the 86 apps, 36\% (72 vs. 53) higher than the best baseline (Stoat).
We also compare the similarities and differences of the bugs between Stoat and our approach, and the results show that its detected bugs are a subset of our detected bugs. 
This indicates the effectiveness of {\tool} in detecting bugs and helps to ensure app quality. 

We can also see that, in every time point, {\tool} detects more bugs than the baselines, and reaches the highest value in about 17 minutes, saving 35\% (17 vs. 26) of the testing time compared with the best baseline (also with more detected bugs). 
This again proves its effectiveness and efficiency of {\tool}, which is valuable for saving more time for the follow-up bug fixing.
We will conduct a further discussion about the reason behind the superior performance in Section \ref{sec_results_RQ3}.


\input{figure/RQ2-result.tex}

\subsubsection{\textbf{Performance of Operation Matching (RQ3)}}
\label{sec_results_RQ3}

When testing the experimental apps in RQ1, we randomly choose 1,000 LLM's feedback answer, i.e., natural language described operations outputted by LLM, and evaluate whether the operations can successfully match the correct GUI widgets.
Specifically, the two authors follow the principle of open-coding, analyze the feedback answer and find the matching widget on the current page. 
For the inconsistent labeling, the third author will judge until all the authors reach an agreement.

Results show that the operation matching can achieve 0.96 accuracy, indicating that most of the LLM's feedback answers can be accurately matched to the GUI widget. 
This lays a solid foundation for the high activity coverage of our {\tool}.



\input{figure/Case-study.tex}

%% file: tab/dataset.tex
\begin{table}[h]
\vspace{-0.05in}
\caption{Dataset of effectiveness evaluation.}
\vspace{-0.05in}
\label{tab:app-info}
\centering
\footnotesize
\begin{tabular}{p{1.2cm}<{\centering} | p{1.2cm}<{\centering}| p{1.2cm}<{\centering} | p{1.2cm}<{\centering} | p{1.2cm}<{\centering}}
\toprule
\textbf{Statistics} & \textbf{\#Activities}  & \textbf{\#Bugs} & \textbf{\#Download} & \textbf{\#Update} \\ 
\midrule
\textbf{Min} & 7 & 1 & 50K+ & 05/22 \\
\textbf{Max} & 21 & 9 & 50M+ & 11/22 \\
\textbf{Median} & 10 & 3 & 1M+ & - \\
\textbf{Average} & 9 & 1.5 & 10M+ & - \\
\midrule
\textbf{All} & 790 & 129 & - & - \\
\bottomrule
\end{tabular}
\vspace{-0.05in}
\end{table}


%% file: tab/RQ1-result.tex
\begin{table*}[h]
\vspace{0.15in}
\caption{Performance of activity coverage (RQ1).}
\vspace{-0.05in}
\label{tab:RQ1-result}
\centering
\footnotesize
\begin{tabular}{p{2.7cm}<{\centering} || p{1.2cm}<{\centering}  | p{1.0cm}<{\centering} || p{0.6cm}<{\centering} | p{0.6cm}<{\centering} | p{1.2cm}<{\centering} | p{1.2cm}<{\centering} | p{1.3cm}<{\centering} || p{1.2cm}<{\centering} | p{1.2cm}<{\centering} || p{1.2cm}<{\centering} }
\toprule
\multirow{2}*{\textbf{Metric}} & \multicolumn{2}{c||}{\textbf{Random-/rule-based}} & \multicolumn{5}{c||}{\textbf{Model-based}} & \multicolumn{3}{c}{\textbf{Learning-based}} \\
 & Monkey & Droidbot & Stoat & Ape & Fastbot & ComboDroid & TimeMachine & Humanoid & Q-testing & \textbf{{\tool}} \\ 
\midrule
\textbf{\#Widgets} & 351 & 893 & 1337 & 1582 & 1437 & 1388 & 1701 & 1453 & 1398 & \textbf{1989} \\
\textbf{\#Activities} & 104 & 269 & 333 & 370 & 391 & 383 & 401 & 340 & 323 & \textbf{477} \\
\textbf{Avg. activity coverage} & 0.16 & 0.34 & 0.44 & 0.51 & 0.56 & 0.53 & 0.57 & 0.49 & 0.45 & \textbf{0.71} \\
\bottomrule
\end{tabular}
\vspace{-0.05in}
\end{table*}


%% file: figure/RQ1-ActivityCoverage.tex
\begin{figure}[htb]
\centering
\vspace{-0.05in}
\includegraphics[width=8.4cm]{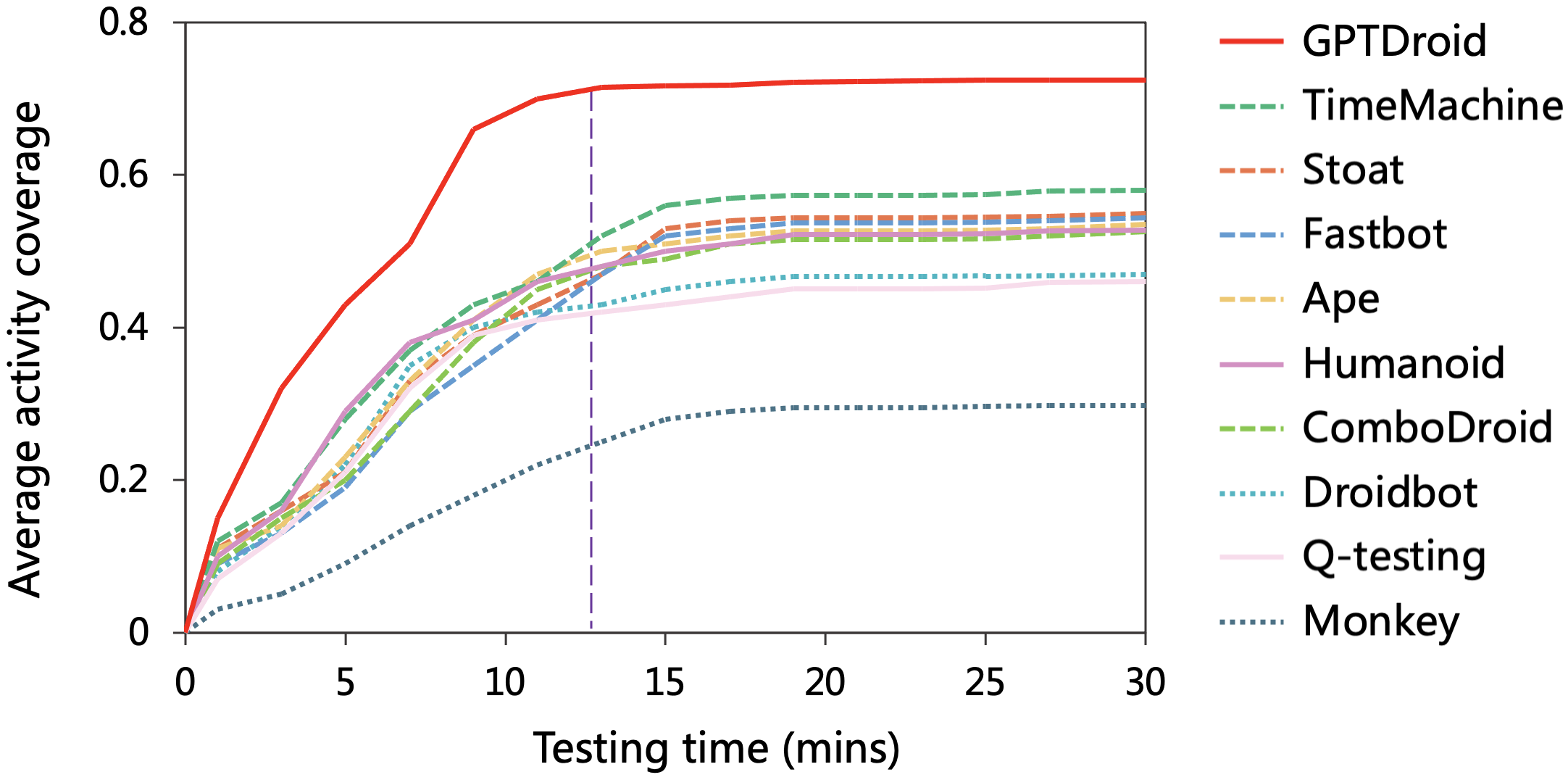}
\vspace{-0.05in}
\caption{Activity coverage with varying time (RQ1).}
\label{fig:RQ1-ActivityCoverage}
\vspace{-0.05in}
\end{figure}

%% file: figure/RQ2-result.tex
\begin{figure}[htb]
\centering
\vspace{-0.05in}
\includegraphics[width=8.3cm]{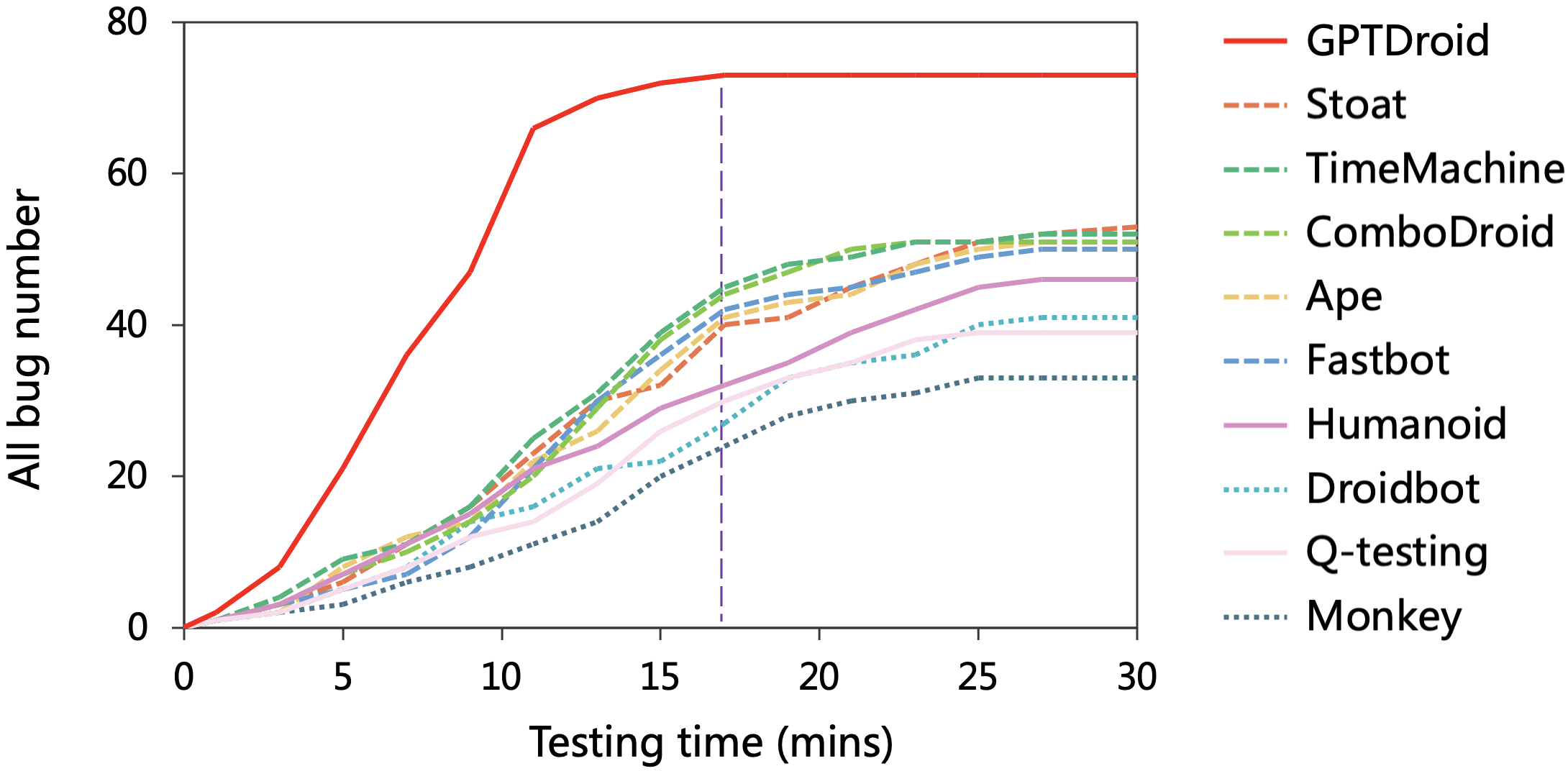}
\vspace{-0.05in}
\caption{Bug detection with varying time (RQ2).}
\label{fig:RQ2-result}
\vspace{-0.1in}
\end{figure}

%% file: figure/Case-study.tex
\begin{figure*}[t]
\centering
\vspace{0.05in}
\includegraphics[width=17.2cm]{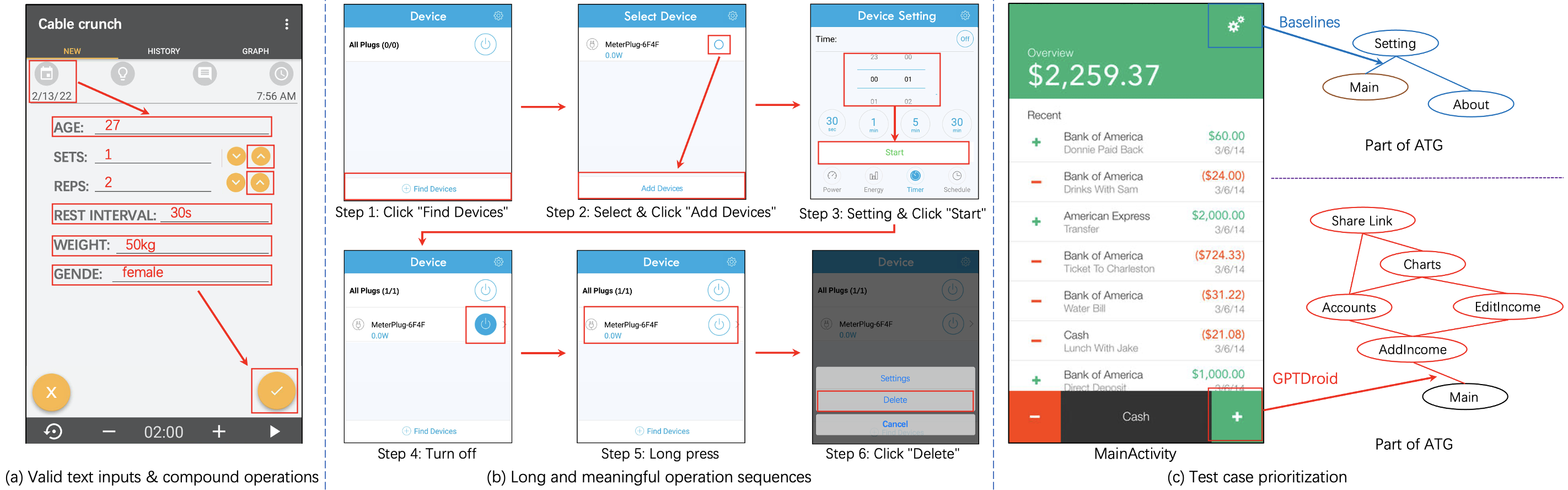}
\vspace{-0.05in}
\caption{Examples of our finding.}
\label{fig:case-study}
\vspace{-0.1in}
\end{figure*}

%% file: sec/usefulness.tex
\section{Usefulness Evaluation}
\label{sec_Usefulness}

\subsection{Experimental Setup}
\label{sub_Usefulness_Experimental_Setup}

This section further evaluates the usefulness of {\tool} in detecting new crash bugs.
We employ a similar experimental setup to the previous section. 
To make it brief, we only compare the best baselines for bug detection in the last section, i.e., Droidbot, Stoat and Humanoid, the best one from each type of methods (random-/rule-based methods, model-based methods and learning-based methods).

We begin with the most popular and recently updated 317 apps from 12 categories as in the previous section. 
Then we reuse the five criteria in the previous section for filtering the unusable apps.
Differently, we loosen the criteria 5, which only requires the app to have ways for bug reporting, since the issue records or pull requests are not mandatory in this section.
This results in 216 apps for our usefulness evaluation. 
Note that, this section aims at evaluating whether {\tool} can detect new bugs on these apps, thus the overlap between the apps of this section and the previous section is allowed.

We use the same hardware and the software configurations as the previous evaluation section. 
When the crash bugs are detected, we report them to the app development team through online issue reports or email.

\input{tab/usefulness.tex}

\subsection{\textbf{Results and Analysis}}
\label{subsub_Effectiveness_Result}

For the 216 apps, {\tool} detects 135 bugs involving 115 apps, of which 48 bugs involving 39 apps are newly-detected bugs. 
Furthermore, these new bugs are not detected by the three baselines. 
We submit these 48 bugs to developers, and 25 of them have been fixed/confirmed so far (8 fixed and 17 confirmed), while the remaining are still pending (none of them is rejected). 
 This further indicates the effectiveness of our proposed {\tool} in bug detection.
Due to space limit, Table \ref{tab:RQ3-Usefulness} presents these fixed/confirmed bugs, and the full lists can be found on our website\textsuperscript{\ref{github}}.

We further analyze the details of our found bugs, and 17 of them involve multiple text inputs or compound operations. 
Besides, we also observe that there are 11 bugs with more than 20 operation steps before triggering the bug, counting from the \textit{MainActivity} page, which indicates the ability of {\tool} in testing deeper features.
Furthermore, we find at least 28 bugs related to the main business logic of the app, for example, a bug about the health data statistics is revealed for a digital health app. 


%% file: tab/usefulness.tex
\begin{table}[h]
\caption{confirmeded or fixed bugs}
\vspace{-0.05in}
\label{tab:RQ3-Usefulness}
\centering
\footnotesize
\begin{tabular}{p{0.55cm}<{\centering} | p{1.85cm}<{\centering} | p{1.2cm}<{\centering} | p{1.2cm}<{\centering} | p{1.4cm}<{\centering}}
\toprule
\textbf{Id} & \textbf{APP Name} & \textbf{Category} & \textbf{Download} & \textbf{Status}\\
\midrule
1 & PerfectPia & Music & 50M+ & confirmed\\

2 & MusicPlayer & Music & 50M+ & confirmed\\  

3 & NoxSecu & Tool  & 10M+ & fixed\\  

4 & Degoo& Tool & 10M+ & fixed\\  

5 & Proxy & Tool & 10M+ & confirmed\\  

6 & Secure & Tool & 10M+  & confirmed\\  

7 & Thunder & Tool & 10M+  & confirmed \\

8 & ApowerMir & Tool & 5M+  & confirmed\\  

9 & MediaFire & Product & 5M+  & confirmed\\  

10 & Postegro & Commun & 500K+ & fixed\\   

11 & Deezer MP & Music & 500K+  & fixed\\  

12 & MTG & Utilities & 500K+ & fixed \\  

13 & OFF & Health & 500K+ & confirmed \\ 

14 & Yucata & Tool & 500K+ & confirmed \\  

15 & ClassySha & Tool & 500K+  & confirmed \\  

16 & Linphone & Commun & 500K+  & confirmed \\ 

17 & Paytm & Finance & 100K+ & confirmed \\

18 & Transdroid & Tool & 100K+ & confirmed \\ 

19 & Transistor & Music  & 10K+ & fixed \\

20 & Onkyo & Music  & 10K+  & fixed \\

21 & Democracy & News  & 10K+ & confirmed \\

22 & NewPipe  & Media  & 10K+ & confirmed \\ 

23 & LessPass & Product  & 10K+ & confirmed \\  

24 & CEToolbox &  Medical  & 10K+  & confirmed \\ 

25 & OSM  & Health  & 10K+  & fixed \\  

\bottomrule

\hline
\end{tabular}
\vspace{-0.05in}
\end{table}

%% file: sec/discussion.tex
\section{Discussion}
\label{Sec_Discussion}
Despite the superior performance of {\tool} in the last section, it is still unclear the reason behind it.
To fully understand the testing capability of the LLM, we carry out a qualitative study by investigating the cases in which our model outperforms baselines.
We summarize four kinds of capabilities including low-level (i.e., valid text input, and compound actions) and high-level ones (i.e., long meaningful test trace, and test case prioritization).
These findings pave the way for further research in this area.


\textbf{Valid text inputs.} Our approach can automatically fill in valid text content to the input widget which is essentially the key for passing the page as seen in Figure \ref{fig:case-study} (a).
More importantly, our model can generate semantic text input (e.g., income, date, ID number, searching item, etc) accordingly.
Besides single text input, it can also successfully fill in multiple input widgets at the same time which are correlated to each other like the departure and arrival cities and dates in the flight booking app.
It may be due to GPT-3's language generation capabilities~\cite{brown2020GPT3} which was well learned during the training phase. 


\textbf{Compound actions.} {\tool} can conduct complex compound operations guided by the LLM. 
As shown in the above example (Figure \ref{fig:case-study} (a)), to add the ``Cable crunch'' information, it first inputs the text, selects the date, sets the ``SETS'' and ``REPS'' by clicking the upper or lower button, then click the submit button in the lower right corner. 
Since there are many tutorials or bug reports on the web including natural-language descriptions of how certain actions can lead to specific outcomes which may be adopted into GPT-3's training corpus.
It may help with a better understanding of the causal relationships between actions and outcomes, resulting in LLM's compound action ability in GUI testing.

\textbf{Long meaningful test trace.}
{\tool} can automatically generate the test cases with a long sequence of operations which together accomplish a business logic of the app, and this is quite important for covering the app features and ensuring its quality. 
As shown in Figure \ref{fig:case-study} (b), in SmartMeter app~\cite{SmartMeter}, to test a commonly-used app feature ``delete equipment'', the automatic tool first needs to click ``find devices'' in the device page, then select a device (Bluetooth is turned on and there are candidate devices) and click ``add device'' for adding it in the device page; input the related information and click ``start'' to start the device; then turn off this device in the device page, long press it and click ``delete'' from the pop-up menu.
Only with this long sequence of operations which touches the ``deleting equipment'' feature, a crash can be revealed.
It may be because of GPT-3's exposure to tutorial or bug reports which contain step-by-step instructions or descriptions of how to trigger a certain feature or reproduce a certain bug \cite{su2021benchmarking,wang2022detecting} in the training corpus.
Therefore, provided with low-level semantic information (i.e., current GUI page) and high-level testing history, {\tool} can capture long-term dependencies among GUI pages for generating long meaningful exploration sequences.

\textbf{Test case prioritization.} 
We also observe that {\tool} usually prioritizes testing the ``important'' widgets, which is valuable for reaching a higher activity coverage and covering more key activities with relatively less time.
As shown in Figure \ref{fig:case-study}(c), in the Main page of the Moni app~\cite{Moni}, the baseline tools tend to first click the ``Setting'' button following the exploration order from upper to lower, which leads the testing easily trapped into the setting page cycle.
Our {\tool} chooses to first test the ``AddIncome'' activity, i.e., click the ``add'' button, which is based on the semantics of the GUI page and app information, and can quickly explore the activities related to the key features of the app. 
This may be because that GPT-3's training corpus contains a wide variety of software-related information, including user manuals, release notes, and app/software descriptions where developers often highlight the most important features at the front.

%% file: sec/related.tex
\section{Related Work}
\label{sec_related}

\subsection{Mobile App GUI and GUI Analysis.} 
GUI provides a visual bridge between apps and users, and is drawing increasing attention~\cite{DBLP:conf/chi/SpeicherBG15,DBLP:conf/chi/NebelingSN13,wei2016taming}.
The graphical user interface (GUI) is the most important type of UI for most mobile apps, where apps present content and actionable widgets on the screen and users interact with the widgets using actions such as clicks, swipes, and text inputs.
GUI is an indispensable part of software on most major platforms including Android. Analyzing the app’s GUI is of great interest to many researchers and practitioners.
Related studies included automatic GUI search~\cite{reiss2018seeking, behrang2018guifetch,chen2019gallery}, GUI code generation~\cite{nguyen2015reverse, chen2018ui, moran2018machine}, GUI changes detection and summarization~\cite{moran2018automated,moran2018detecting}, GUI design~\cite{chen2021should,yang2021don}, etc.
Gao et al.~\cite{gao2017every} and Li et al.~\cite{li2019characterizing} analyzed the possible problems in UI rendering, and developed automatic approaches to detect them.
Nayebi et al.~\cite{nayebi2012state} and Holzinger et al.~\cite{holzinger2012making} found that different resolutions of mobile devices have brought challenges in Android app design and implementation.
Huang et al.~\cite{huang2014asdroid} and Rubin et al.~\cite{rubin2015covert} proposed to detect stealthy behaviors in Android apps by comparing the actual behaviors with the UI. 
Chen et al.~\cite{Chen2020From} introduced a machine learning-based method to extract UI skeletons from UI images, in order to facilitate GUI development.
In human-computer interaction research, software GUI is mainly used to mine UI design practices~\cite{kumar2013webzeitgeist,alharbi2015collect} and interaction patterns~\cite{deka2016erica} at scale. The mined knowledge can further be used to guide UI and UX (user experience) design. 
Due to these challenges, how to reasonably summarize the app is a problem to be solved. Our focus is to use the reasonable natural language to describe the GUI information of the app page, so that the large language model can understand the GUI information.

\subsection{Automated GUI testing}
To ensure the quality of mobile apps, many researchers study the automatic generation of large-scale test scripts to test apps~\cite{xie2007designing}.
Monkey~\cite{Monkey} is the popular random-based automated GUI testing tool, which emits pseudo-random streams of UI events and some system events. 
It is easy to use and compatible with different Android versions.  
However, the random-based testing strategy cannot formulate a reasonable testing path according to the characteristics of the app, resulting in low test coverage. 
To improve the test coverage, researchers propose model-based~\cite{dong2020time,mirzaei2016reducing,yang2018static,yang2013grey,zeng2016automated} automated GUI testing methods, design corresponding models through the research and analysis of large-scale apps.
Sapienz~\cite{mao2016sapienz} used genetic algorithms as the model and Stoat~\cite{su2017guided} used the stochastic model learned from an app to optimize test suite generation. 
Ape~\cite{gu2019practical} used the runtime information to dynamically evolve its abstraction criterion via a decision tree and generated UI events via a random and greedy depth-first state exploration strategy.
ComboDroid~\cite{wang2020combodroid} obtained such use cases either from humans or automatically generates from a GUI model constructed by GUI exploration and analyzed the data flow between obtained use cases, and combined them to generate final tests. 
Although model-based automated GUI testing tools can improve test coverage, the coverage is still low because it does not consider the semantic information of the app's GUI and Page.
Researchers further proposed human-like testing strategies and designed learning-based automated GUI testing methods. 
Humanoid~\cite{li2019humanoid} used a deep neural network model that predicts which UI elements on the current GUI page are more likely to be interacted with by users and how to interact with them.
Q-testing~\cite{pan2020reinforcement} used a reinforcement learning-based method to compare GUI pages and give rewards. These rewards are used and iteratively updated to guide the testing to cover more functionalities of apps.
Although the learning-based approach can improve the test coverage by learning a large number of interactive processes or using the idea of reinforcement learning. However, it is still unable to better understand the semantic information of the page and plan the path according to the actual situation of the app, and is greatly affected by the training data.
This study aims at proposing a more effective approach to generate human-like actions for thoroughly and more effectively testing the app, and accomplishing it with LLM. 

\subsection{Large Language Model}
Recently, there has been a great success of pre-trained Large Language Models (e.g.,  RoBERTa~\cite{2019RoBERTa}, GPT-3~\cite{brown2020GPT3}, PaLM~\cite{chowdhery2022palm},
OPT~\cite{zhang2022opt}) in a variety of NLP tasks. 
Due to the large amounts of available pre-training data from the internet, research shows that LLMs can already be used for very specific downstream tasks through the new paradigm ``pre-train, prompt and prediction''~\cite{liu2023pre} without any fine-tuning of special data sets.
This paradigm for LLM was widely used in many works and achieved state-of-the-art performance on downstream tasks~\cite{huang2022prompt,deng2022fuzzing}. The core of this paradigm is to use prompt engineering~\cite{liu2023pre,Branwen2020Gpt-3creative, Cantino201Prompt,ge2021visual}, where a natural language description of the task is provided to the LLM. 
Considering the powerful performance of LLM, researchers try to use LLM to solve relevant tasks in the field of software engineering.
Supported by code naturalness~\cite{hindle2016naturalness}, researchers applied the LLMs to code writing in different programming languages~\cite{chen2021evaluating,feng2020codebert,fried2022incoder,xu2022systematic}. Huang et al.~\cite{huang2022prompt} used LLM for the type inference in statically-typed partial Code. In testing, 
LLMFuzz~\cite{deng2022fuzzing} used LLMs to generate input programs for fuzzing Deep Learning libraries. 
Xia et al.~\cite{xia2022practical,xia2022less} applied LLM to automatic program repair to improve the accuracy of the generated repair patches.
Austin et al.~\cite{austin2021program} and Jain et al.~\cite{jain2022jigsaw} used LLM for program synthesis in general-purpose programming languages.
This paper opens a new dimension for automated GUI testing by formulating it as a Q\&A task with LLM.

%% file: sec/conclusion.tex
\section{Conclusion}
\label{sec_conclusion}

As one of the most important quality assurance activities for mobile apps, 
automated GUI testing has made much progress, yet still suffers from low activity coverage and may miss critical bugs.
This paper aims at generating human-like actions to facilitate app testing more thoroughly and effectively. 
Inspired by the success of LLM like ChatGPT, we formulate the GUI testing problem as a Q\&A task and propose {\tool}.
It extracts the static and dynamic context of the current GUI page, encodes them into prompt question to ask the LLM, decodes the LLM's feedback answer into actionable operations to execute the app, and iterates the whole process. 
Results on 86 popular apps demonstrate that {\tool} can achieve 71\% activity coverage, with 32\% higher than the best baseline, and can detect 36\% more bugs with faster speed than the best baseline. 
{\tool} also detects 48 new bugs on Google Play with 25 of them being confirmed/fixed, with the remaining pending.
The capability of {\tool} in generating semantic text input and compound actions, guiding to explore the long meaningful test trace, and prioritizing test cases, further proves the effectiveness and human-like aspects of our proposed {\tool}.

In the future, we will fine-tune the LLM to improve the performance and conduct a systematic study to understand reasons why LLM can help the GUI testing. 